\documentclass[prl,twocolumn,aps,showpacs,nofootinbib,nobibnotes,superscriptaddress]{revtex4}

\usepackage{epsfig}
\usepackage{amssymb}
\usepackage{graphics}
\usepackage{bm}

\begin{document}

\draft

\title{Gravitational Radiation from Primordial Helical MHD Turbulence}

\date{\today}

\date{\today~~KSUPT-08/1}
\author{Tina Kahniashvili}
\affiliation{CCPP, New York University, 4 Washington Plaza, New
York,  NY 10003, USA}\affiliation{Department of Physics, Kansas
State University, 116 Cardwell Hall, Manhattan, KS 66506, USA}
\affiliation{Department of Physics, Laurentian University, Ramsey
Lake Road, Sudbury, ON P3E 2C6, Canada} \affiliation{National
Abastumani Astrophysical Observatory, 2A Kazbegi Ave, Tbilisi,
GE-0160, Georgia}
\author{Grigol Gogoberidze}
\affiliation{Centre for Plasma Astrophysics, K.U.\ Leuven,
Celestijnenlaan 200B, 3001 Leuven,
Belgium}\affiliation{Department of Physics, Kansas State
University, 116 Cardwell Hall, Manhattan, KS 66506,
USA}\affiliation{National Abastumani Astrophysical Observatory,
2A Kazbegi Ave, Tbilisi, GE-0160, Georgia}
\author{Bharat Ratra}
\affiliation{Department of Physics,
Kansas State University, 116 Cardwell Hall, Manhattan, KS 66506,
USA}

\begin{abstract}
We consider gravitational waves (GWs) generated by primordial
inverse-cascade helical magneto-hydrodynamical (MHD) turbulence
produced by bubble collisions at the electroweak phase transitions
(EWPT). Compared to the unmagnetized EWPT case, the spectrum of
MHD-turbulence-generated GWs peaks at lower frequency with larger
amplitude and can be detected by the proposed Laser Interferometer
Space Antenna (LISA).
\end{abstract}


\maketitle


When detected, primordial cosmological GWs will provide a  very
valuable probe of the very early Universe \cite{sources}. Various
mechanisms that generate such GWs have been discussed: quantum
fluctuations \cite{inflation};
  bubble wall motion
and collisions during phase transitions \cite{bubble,kos1};
cosmic strings \cite{strings};
 cosmological magnetic fields \cite{magnet,cdk04,cd06};
 and plasma turbulence
\cite{kmk02,dolgov,kgr05,gkk07}. From the direct detection point
of view, GWs generated during the EWPT are promising since their
peak frequency lies in or near the  LISA \cite{lisa} frequency
band \cite{m00}, however, to produce a detectable signal the EWPT
must be strong enough \cite{detection,gs06,grojean}. Currently
discussed EWPT models do not predict an observable GW signal from
bubble collisions \cite{grojean}, nor for GWs produced by
unmagnetized turbulence \cite{gkk07}.

Here we study the generation of GWs during a first-order EWPT
assuming that bubble collisions produce helical MHD
turbulence.\footnote{Kinetic or magnetic helicity generation at
the EWPT is studied in Refs. \cite{helicity}. Previously we
studied generation of GWs by direct-cascade  turbulence and found
that, due to parity violation in the early Universe, the induced
GWs are circularly polarized \cite{kgr05}. Polarized GWs are
present in other models \cite{gw-pol}, and the polarization of
the GW background is in principle observable, either directly
\cite{seto} or through  the CMB \cite{cdk04,a}.} In the case of
unmagnetized hydrodynamical turbulence the peak frequency of the
GW power spectrum is determined by the inverse turn-over time of
the largest  eddy and the energy-scale when the GW is generated.
Recently discussed modifications of the standard EWPT model place
the transition at a
 higher energy-scale \cite{gs06}. As a result, the GW power
spectrum peak frequency is shifted to  higher frequency which,
since the GW spectrum is sharply peaked, reduces the possibility
of detection by LISA. On the other hand, in the case of MHD
turbulence the presence of an energy inverse-cascade leads to an
increase in the effective size of the largest eddy (now
associated with an helical magnetic field), and can result in the
GW power spectrum peaking in the LISA band, with amplitude large
enough to be detected by LISA. We adapt the technique developed
in Ref. \cite{gkk07} to study this case here.   We model MHD
turbulence and  obtain the GW spectrum by using an analogy with
the theory of sound wave production by hydrodynamical turbulence
\cite{L52,P52,G,my75}.


Since the turbulent fluctuations are stochastic, so are the
generated GWs. The GW energy density  is \cite{m00}
\begin{eqnarray}
\rho_{\rm GW} ({\bf x}) &=& \frac{1}{32\pi G} \langle
\partial_t h_{ij}({\mathbf x},t) \partial_t h_{ij} ({\mathbf
x},t)\rangle
\label{eq:04} \\
&=&\frac{G}{2\pi } \int {\rm d}^3 {\bf x}^\prime {\rm d}^3 {\bf
x}^{\prime \prime} \frac{\langle \partial_t S_{ij}({\mathbf
x}^\prime,t^\prime) \partial_t S_{ij}({\mathbf x}^{\prime
\prime},t^{\prime \prime}) \rangle }{|{\bf x}- {\bf x}^\prime|
|{\bf x}- {\bf x}^{\prime \prime}|}. \nonumber
\end{eqnarray}
Here the times $t^{\prime (\prime\prime)}=t-|{\bf x}- {\bf
x}^{\prime (\prime\prime)}|$, $i$ and $j$ are spatial indices
(repeated indices are summed),  the source $S_{ij}({\mathbf x},
t)= T_{ij} ({\mathbf x}, t) - \delta_{ij} T^k_k ({\mathbf x}, t)
/3$ is the traceless part of the stress-energy tensor $T_{ij}$,
$G$ is the gravitational constant, and we use natural units with
$\hbar = 1 = c$. We assume that the turbulence exists for a time
short enough to neglect the cosmological expansion during GW
production. We consider metric perturbations in the far-field
limit (i.e.\ for $x \gg d$, where $d$ is a characteristic
length-scale of the source region), where GWs are the only metric
perturbations \cite{W}, and  replace $|{\bf x}-{\bf x}^\prime|$
by $|\bf x|$ in Eq. (\ref{eq:04}). If the turbulence is
stationary then the GW spectral energy density $I({\bf
x},\omega)$ ($\rho_{\rm GW} ({\bf x}) = \int {\rm d} \omega
I({\bf x},\omega)$ where $\omega$ is the angular frequency) is
\cite{gkk07}
\begin{equation}
I({\bf x},\omega) =\frac{4\pi^2\omega^2 G {\rm w}^2}{|{\bf x}|^2}
\int {\rm d}^3 {\bf x}^\prime H_{ijij} \left( {\bf x}^\prime,
\frac{{\bf x}}{|{\bf x}|} \omega, \omega\right). \label{eq:20}
\end{equation}
Here $H_{ijij} ( {\bf x}^\prime, {\bf k}, \omega )$ (where ${\bf
k} $ is a proper wavevector)  is the (double traced)
four-dimensional Fourier transform of the two-point time-delayed
fourth-order correlation tensor, $ \langle S_{ij}({\bf x}^\prime
,t) S_{lm}({\bf x}^{\prime \prime}, t+\tau) \rangle/{\rm w}^2$,
with respect to ${\bf x}^{\prime \prime}-{\bf x}^{\prime}$ and
$\tau$, where ${\rm w}=\rho+p$ is the enthalpy density and $p$
and $\rho$ the pressure and energy density of the plasma.

We assume that primordial MHD turbulence is generated at time
$t_\star$  at proper length-scale $l_0=2\pi/k_0$ with
characteristic velocity perturbation $v_0$.\footnote{We assume
that the usual and magnetic Reynolds numbers are much greater
than unity on scales $\sim l_0$, otherwise there is no
turbulence. Throughout this paper the symbol $\sim$ represents
equality to the accuracy of a dimensionless multiplicative
constant of order unity.} The dynamics of MHD turbulence is
dominated by Alfv\'en waves for which the magnetic and kinetic
energy densities are in approximate equipartition \cite{B03}. In
this case $v_0 \sim b_0$, where $b_0 = B_0/\sqrt{4\pi {\rm w}}$
is the characteristic magnetic field perturbation expressed in
velocity units. While MHD turbulence is isotropic on large
scales, it is locally anisotropic on small scales \cite{s83},
resulting in small-scale anisotropy in the generated GW
background. However, GWs are generated mainly by the largest
eddies \cite{gkk07} so we adopt an isotropic turbulence model,
and thus the magnetic field two-point correlation function is
$\langle b_i^\star ({\bf k}, t) b_j({\bf k'}, t+\tau) \rangle =
F_{ij}^M\!({\bf k},t)f(\eta(k),\tau) \delta({\bf k} -{\bf k'})$,
with \cite{my75}
\begin{equation}
F_{ij}^M\!({\bf k},\tau) =  P_{ij}({\bf k})
\frac{E^M\!(k,t)}{4\pi k^2}  + i \varepsilon_{ijl} {k_l}
\frac{H^M\!(k,t)}{8\pi k^2} . \label{eq:4.1}
\end{equation}
Here  $P_{ij}({\bf k}) = \delta_{ij}-{k_i k_j}/{k^2}$, and
$E^M\!(k,t)$ and $H^M\!(k,t)$ are the magnetic field energy and
helicity densities. The Schawarz inequality implies $|H^M\!(k,t)|
\leq 2 E^M\!(k,t)/k $ \cite{B03}. For the total magnetic energy
${\mathcal E}_M(t)=\int E^M\!(k,t) dk$ and helicity ${\mathcal
H}_M(t)=\int H^M\!(k,t) dk$ we get ${\mathcal H}_M(t) \leq 2
\xi_M(t) {\mathcal E}_M(t)$, where $\xi_M(t) \equiv \int
E^M(k,t)k^{-1} dk/{\mathcal E}_M(t)$ is the magnetic-eddy
correlation length. $\eta(k)$ is an autocorrelation function that
determines the characteristic function $f(\eta(k),\tau)$ that
describes the temporal decorrelation of turbulent fluctuations.
In the following we use $f(\eta (k),\tau)=\exp \left( -{\pi}
\eta^2(k) \tau^2/4 \right)$ \cite{K64}.

After generation primordial turbulence freely decays. We adopt the
decaying MHD turbulence model of Refs.\ \cite{BM99,CHB05}. For
non-zero  initial  magnetic helicity turbulence decay is a two
stage process. First decay stage dynamics is governed by a direct
cascade of energy density lasting for a time $\tau_{s0} =s_0
\tau_0$, several times ($s_0 \sim 3-5$) longer then the
characteristic largest-eddy turn-over time $\tau_0=l_0/ v_0
=2\pi/k_0 v_0$. During the first stage energy density flows from
large to small scales and finally dissipates on scales $\sim
l_d=2\pi/k_d$ ($k_d \gg k_0$) where one of the Reynolds numbers
becomes of order unity. Due to the selective decay effect
\cite{B03} magnetic helicity is nearly conserved during this
stage \cite{CHB05}. To compute  the GWs generated by decaying MHD
turbulence, we assume that decaying turbulence lasting for time
$\tau_{s0}$ is equivalent to stationary turbulence lasting for
time $\tau_{s0}/2$. This can be justified using the Proudman
\cite{P52, my75} argument for (unmagnetized) hydrodynamical
turbulence. Consequently, when computing  the emitted GWs we
ignore the time dependence of $E^M(k,t)$ and $H^M(k,t)$. We also
assume small initial magnetic helicity, $\alpha_\star \equiv
{\mathcal H}_M(t_\star)/ [2 \xi_M(t_\star) {\mathcal
E}_M(t_\star)] \ll 1$. For $E^M(k,t)$ and $\eta(k)$ we use the
Kolmogorov model,
\begin{equation}
E^M\!(k,t) \sim  \varepsilon^{2/3} k^{-5/3},~~~\eta (k)=
\varepsilon^{1/3} k^{2/3}/\sqrt{2\pi}, \label{eq:4.2}
\end{equation}
for $k_0<k<k_d$. Here $\varepsilon \sim k_0v_0^3$ is the energy
dissipation rate per unit enthalpy.

At the end of the first stage turbulence relaxes to a maximally
helical state, $\alpha_{s0} \sim 1$ \cite{CHB05,jedamzik}.
Accounting for conservation of magnetic helicity, the
characteristic velocity and  magnetic field perturbations at this
stage are $v_1 \sim \alpha_\star^{1/2}v_0$ and $b_1 \sim
\alpha_\star^{1/2}b_0$. Second stage dynamics is governed by a
magnetic helicity inverse cascade. If both Reynolds numbers are
large at the end of the first stage, magnetic helicity is
conserved during the second stage. The magnetic eddy correlation
length evolves as $\xi_M(t) \sim l_0\sqrt{1+t/\tau_1}$
\cite{BM99,CHB05} where $\tau_1 \sim l_0/v_1 =
\tau_0/\sqrt{\alpha_\star}$ is the characteristic energy
containing eddy turn-over time at the beginning of the second
stage. The magnetic ${\mathcal E}_M(t)$ and kinetic ${\mathcal
E}_K(t)$ energy densities evolve as \cite{BM99,CHB05}
\begin{eqnarray}
{\mathcal E}_M(t) &\propto &(1+t/\tau_1)^{-1/2} {\rm w} b_1^2,
\nonumber
\\
{\mathcal E}_K(t) &\propto &(1+t/\tau_1)^{-1} {\rm w} v_1^2.
\label{eq:4.06}
\end{eqnarray}
These imply that the characteristic turn-over ($\tau_{\rm to}$)
and cascade ($\tau_{\rm cas}$) timescales evolve as {
\begin{eqnarray}
\tau_{\rm to} \sim \tau_{\rm cas} \sim \tau_1 (1+t/\tau_1). \label{eq:4.7}
\end{eqnarray}

To compute the  GWs emitted during the second stage we  use the
stationary turbulence model that has the same GW output.
Introducing the characteristic wavenumber $k_\xi(t) =
2\pi/\xi_M(t)$ and using Eqs.\ (\ref{eq:4.06}) we find ${\mathcal
E}_M \sim {\rm w} v_1^2 k_\xi(t)/k_0$ and ${\mathcal E}_K \sim
{\rm w} v_1^2 [k_\xi(t)/k_0]^2$ since $b_1 \sim v_1$. The time
when turbulence is present on  scale $\xi_M(t)$ is determined by
Eq.\ (\ref{eq:4.7}) which can be rewritten as $\tau_{\rm cas} \sim
\tau_1[k_0/k_\xi(t)]^2$. So instead of considering decaying
turbulence, we  consider stationary turbulence with a
scale-dependent duration time (time during which the magnetic
energy is present at the scale), $\tau_{s1} \sim \tau_1[k_0/k]^2$
(for $k =k_\xi$ this  coincides with $\tau_{\rm cas}$).

The expression for  ${\mathcal E}_M$  yields the time-independent
\begin{equation}
E^M\!(k,t) = C_{1} v_1^2/k_0= k H^M\!(k,t)/2,~~~
k_S<k<k_0. \label{eq:4.9}
\end{equation}
Here $C_1$ is a constant of order unity, $k_S$ is the smallest
wavenumber where the inverse cascade stops, and the second
equation follows from saturating the causality condition. For the
second stage  autocorrelation function, which is inversely
proportional to the turn-over time (\ref{eq:4.7}), we assume
$\eta(k)=(k/k_0)^2/ \sqrt{2 \pi} \tau_1$. At the largest scales
there is no efficient dissipation mechanism, so the inverse
cascade will be stopped at  scale $l_S(t) = 2\pi/k_S$ where
either the cascade timescale $\tau_{\rm cas}$ reaches the
expansion timescale $H^{-1}_\star=H^{-1}(t_\star)$, or when the
characteristic length scale $\xi_M(t) \sim l_S$ reaches the
Hubble radius. These conditions are $\alpha_\star^{-1/2}l_S^2/v_0
l_0 \leq H^{-1}_\star$ or $l_S \leq H^{-1} _\star$ (the cascade
time is scale dependent and maximal  at $k=k_S$). Defining
$\gamma=l_0/ H^{-1}_\star$ ( $\gamma \leq 1$), it is easy to see
that the first condition is fulfilled first and consequently
$k_0/k_S \leq (v_0 /\gamma)^{1/2} \alpha_\star^{1/4}$. To have an
inverse cascade requires $k_0/k_S \geq 1$, leading to a
constraint on initial helicity, $\gamma \leq M
\alpha_\star^{1/2}$ (where $M=v_0$ is the turbulence Mach number).

The magnetic field perturbation stress-energy tensor is $T_{ij}^M
({\mathbf x}, t)={\rm w} b_i({\mathbf x}, t) b_j({\mathbf x},
t)$. For the first decay stage  we compute  for this magnetic
part and then double the  result  to account for approximate
magnetic and kinetic energy equipartition for Alfv\'en waves.
During the second stage, according to Eqs.\ (\ref{eq:4.06}),
kinetic energy can be neglected compared to magnetic energy. To
compute $ H_{ijij}({\bf k},\omega)$ we assume Millionshchikov
quasi-normality  \cite{my75} and use the convolution theorem (for
details see Sec. III of Ref. \cite{gkk07}).
 Using the ($k \rightarrow 0$) aero-acoustic
approximation, which is accurate for low Mach number ($M \leq 1 $,
(and slightly overestimates GWs amplitude for the Mach number
approaching unity ($M \rightarrow 1$) \cite{gkk07}), we find
\begin{eqnarray}
&&\!\!\!\!\!\!\!H_{ijij}({\bf k}, \omega) \approx H_{ijij}(0,\omega) = ~~~~~~~~~\label{eq:4.08} \\
&&\!\!\!\!\!\!\!\!\frac{7  C_1^2 M^3  \alpha_\star^{3/2}\!} {6
\pi^{3/2} k_0}\!\! \int_{k_S}^{k_0}\!\! \frac{{\rm d}k}{ k^{4}}
\exp\!\left( -\frac{\omega^2 k_0^2}{\alpha_\star M^2 k^4}
\right)\!{\rm erfc} \!\left( - \frac{\omega
k_0}{\sqrt{\alpha_\star} M k^2} \right).
\nonumber
\end{eqnarray}
 The integral  is dominated by the contribution of large scale
 ($k \simeq k_S$) perturbations and is maximal at
$\omega_{\rm max}^{(II)} \sim \alpha_\star^{1/2} M k_S^2/k_0= 2\pi
H_\star$. For the first-stage direct-cascade turbulence the peak
frequency is  $\omega_{\rm max}^{(I)} \sim k_0 M$ \cite{gkk07}.
To determine the peak frequency at the current epoch we need to
account for the cosmological expansion which decreases the GW
amplitude and frequency by the factor $a_\star/a_0$, where
$a_\star$ and $a_0$ are the values of the cosmological scale
factor at the GW generation and current epochs

The total GW energy spectrum at a given space-time event is
obtained by integrating over all source regions with a light-like
separation from that event, and includes contributions from GW
generated during the first and second stages. For  the first
stage (with duration time $\tau^{(I)}_T = s_0\tau_0$)
$\rho_{GW}^{(I)}(\omega)$ is given by Eqs.\ (21) and (A3) of Ref.
\cite{gkk07}. For  the second stage contribution we must account
for the scale dependence of the cascade time. The total GW
fractional energy density parameter at the moment of
emission $\Omega_{{\rm GW}, \star}$ 
is $105~H_\star^4 \omega^3 \sum_m \tau_T^{(m)} H_{ijij}^{(m)}(0,
\omega_\star)/H_0^2 $ \cite{gkk07}. Here the index $m$ runs over
$I$ and $II$ for the first and second decay stages, $H_0$
$\omega_\star$ is an angular frequency at the moment of emission.
The current GW amplitude is related to the current fractional
energy density parameter through $h_C(f) = 1.26 \times 10^{-18}
\left({1\,{\rm
Hz}}/{f}\right)\left[h_0^2\Omega_{GW}(f)\right]^{1/2}$ (where
$h_0$ is the current Hubble parameter $H_0$ in units of 100 ${\rm
km}\,{\rm sec}^{-1}{\rm Mpc}^{-1}$) \cite{m00}, and
\begin{eqnarray}
h_C(f) &\simeq &   2 \times 10^{-14} \left(\frac{100\,{\rm GeV}
}{T_*} \right) \left(\frac{100}{g_*}\right)^{1/3}
 \nonumber\\
&&\times \sum_m \left[\tau_T^{(m)} \omega_\star H_\star^4
H_{ijij}^{(m)}(0, \omega_\star)\right]^{1/2}.~~~~~~~~~~
\label{hctoday}
\end{eqnarray}
Here  the linear frequency $f= (a_\star/a_0) f_\star$ with
$f_\star = \omega_\star/2\pi$, $T_\star$ and $g_\star$ are the
temperature and effective number of relativistic (all fields)
degrees of freedom at scale factor $a_\star$.

\begin{figure}
\includegraphics[width=2.4in]{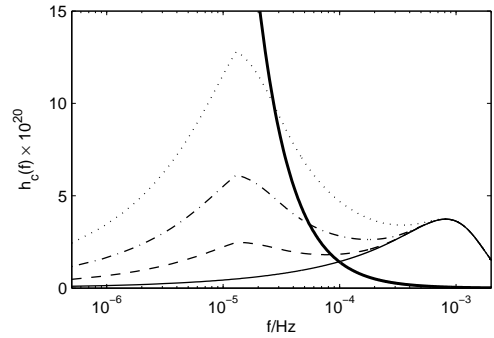}
\caption{The spectrum of gravitational radiation from MHD
turbulence for $g_\ast=100,~T_\ast=100~{\rm GeV},~\gamma=0.01$,
and $M=1/\sqrt{3}$, for four different initial magnetic helicity
values, $\alpha_\star=0$ (solid line), $\alpha_\star=0.02$
(dashed line), $\alpha_\star=0.05$ (dash-dotted line), and
$\alpha_\star=0.1$ (dotted line). The bold   line is the LISA
design sensitivity curve.} \label{fig:spectrum}
\end{figure}

Figure 1 shows $h_C(f)$  for a few  initial magnetic helicity
values. GWs emitted during direct-cascade unmagnetized turbulence
peak at current  $f_{\rm max}^{(I)} \simeq  M \nu_\star$
\cite{gkk07}. We find that the  MHD-inverse-cascade generated GW
(current epoch)  peak frequency is determined by cosmology
parameters, $f_{\rm max}^{(II)} = H_\star a_0/a_\star = 1.6 \times
10^{-5}\,{\rm Hz}\, \left({g_*}/{100}\right)^{1/6}
\left({T_*}/{100\,{\rm GeV}}\right)$ and is independent of
turbulence  parameters. On the other hand, $f_{\rm max}^{(II)}=
\gamma f_{\rm max}^{(I)} /M$  is shifted to lower frequency
compared to the unmagnetized case. From Eq.\ (\ref{hctoday}),
the  amplitude of MHD-turbulence-generated GWs at the peak is a
factor $\sim  \alpha_\star^{9/8} \gamma^{-3/4} M^{3/4}$ larger
than that in the unmagnetized case.

When modeling turbulence  we used the Biskamp and Muller model,
\cite{BM99,CHB05}.  If we adopt the helical MHD turbulence model
of Banerjee and Jedamzik \cite{jedamzik} (also see Refs.
\cite{campanelli}) the GW peak frequency remains the same while
the amplitude of the signal doubles.

Figure 1 shows that even for small values of magnetic helicity the
main contribution to the GW energy density is from the second,
inverse-cascade stage. The GWs will be strongly polarized since
magnetic helicity is maximal at the end of the first stage
\cite{kgr05}. LISA should be able  to detect such GW polarization
\cite{seto}. Unlike the unmagnetized case due to the second
(inverse-cascade) stage contribution the GW amplitude is large
enough at $10^{-4}$ Hz to be detectable by LISA. If the EWPT
occurs at higher energies ($T_\star>100$ Gev) the peak  is
shifted to higher frequency, closer to LISA sensitivity  peak,
which leads to a stronger signal. Our formalism is applicable for
GW production at an earlier QCD phase transition, assuming the
presence of colored magnetic fields \cite{QCD}, or for any other
phase transitions \cite{bubble}; the peak frequency will be
shifted according to the changes in $T_\star$ and $g_\star$. The
GW signal estimated here exceeds that from bubble collisions
\cite{kos1,detection,gs06} or from hydrodynamical, unmagnetized
turbulence \cite{kmk02,dolgov,gkk07}. Of course, this strong
signal assumes   initial non-zero (although small)
 magnetic helicity, so detection of polarized GWs by LISA
 will indicate  parity violation during the EWPT as proposed in
 Refs. \cite{helicity}.

We greatly appreciate useful comments  from A. Gruzinov and the
referees. We acknowledge helpful discussions with R. Durrer, G.
Gabadadze, and A. Kosowsky. T. K. acknowledges the hospitality of
the Abdus Salam International Center for Theoretical Physics. G.
G. and T. K. acknowledge partial support from INTAS
061000017-9258 and Georgian NSF ST06/4-096 grants. B. R.
acknowledges US DOE grant DE-FG03-99EP4103.



\end{document}